\def\complexNumbers{\mathbb{C}}
\def\realNumbers{\mathbb{R}}
\def\integers{\mathbb{Z}}

\def\classicalMSEOperator[#1]{{{\text{MSE}}}\left(#1\right)}
\def\MSEOperator[#1]{{{\text{BMSE}}}\left(#1\right)}
\def\NMSEOperator[#1]{{\text{NBMSE}}\left(#1\right)}
\def\expectationOperator[#1][#2]{{\mathbb{E}_{#2}}\left[#1\right]}
\def\indicatorFunction[#1]{\mathbb{I}\left[{#1}\right]}
\def\probability[#1]{\textrm{Pr}\left({#1}\right)}
\def\complexGaussian[#1][#2]{\mathcal{CN}({#1,#2})}
\def\binomDist[#1][#2]{\mathcal{B}({#1,#2})}
\def\gammaDist[#1][#2]{\Gamma({#1,#2})}
\def\varianceOperator[#1]{\textrm{var}\left(#1\right)}
\def\stdOperator[#1]{\textrm{std}\left[#1\right]}
\def\numberOfEdgeDevices{K}
\def\timeDomainOFDM[#1]{s(#1)}

\def\indexSubcarrier{l}
\def\numberOfActiveSubcarriers{M}

\def\dataSymbols[#1]{d_{#1}}

\def\receivedSymbolAtSubcarrier[#1]{\textbf{r}_{#1}^{(\indexCommunicationRound)}}
\def\transmittedSymbolAtSubcarrier[#1]{t_{#1}^{(\indexCommunicationRound)}}
\def\randomSymbolAtSubcarrier[#1]{s_{#1}^{(\indexCommunicationRound)}}
\def\channelAtSubcarrier[#1]{\textbf{h}_{#1}^{(\indexCommunicationRound)}}
\def\noiseAtSubcarrier[#1]{\textbf{n}_{#1}^{(\indexCommunicationRound)}}
\def\numberOfOFDMSymbols{S}
\def\indexOFDMSymbol{m}
\def\asymbolFromED[#1]{d_{#1}}

\def\exponentialIntegral[#1]{{\rm Ei}(#1)}
\def\tciFactor[#1]{p_{#1}}
\def\mappingFunction{\mathcal{M}}
\def\symbolEnergy{E_{\rm s}}
\def\indexActiveSymbol{\ell}
\def\voteInTime[#1][#2]{m_{#1,#2}}
\def\voteInFrequency[#1][#2]{l_{#1,#2}}
\def\numberOFEDsForOptionGeneral[\indexActiveSymbol]{\hat{U}^{+}_{\indexGradient,\indexDigit}}

\def\vectorX{\textbf{\textrm{x}}}

\def\numberOFEDsForOptionGeneral[#1]{U_{\indexGradient,\indexDigit,#1}}
\def\numberOFEDsForOptionGeneralDetector[#1]{\hat{U}_{\indexGradient,\indexDigit,#1}}
\def\numberOFEDsForOptionGeneralVar[#1]{{U}_{#1}}

\def\noiseVariance{\sigma_{\rm n}^2}

\def\coefficientOne{\sigma_{\rm channel}^2}

\def\correctDecision[#1]{p_{#1}}
\def\incorrectDecision[#1]{q_{#1}}
\def\aparameterForBer[#1]{\epsilon_{#1}}

\def\probabilityIncorrect[#1]{P^{\rm err}_{#1}}

\def\oneVector[#1]{\textbf{\textrm{1}}_{#1}}
\def\zeroVector[#1]{\textbf{\textrm{0}}_{#1}}
\def\identityMatrix[#1]{{\textbf{\textrm{I}}_{#1}}}

\def\dataset[#1]{\mathcal{D}_{#1}}

\def\datasetBatch[#1]{\mathcal{\tilde{D}}_{#1}}
\def\batchSize{n_{\rm b}}

\def\completeData{\mathcal{D}}
\def\numberOfModelParameters{Q}
\def\sampleData[#1]{{\textrm{\textbf{x}}}_{#1}}
\def\sampleLabel[#1]{{y}_{#1}}
\def\learningRate{\eta}

\def\deltaVectorAtIteration[#1][#2]{{\Delta}^{(#1)}_{#2}}
\def\deltaVectorAtIterationEle[#1]{{\bm \Delta}^{#1}}

\def\indexED{k}
\def\indexGradient{q}
\def\indexSampleData{{\ell}}
\def\indexCommunicationRound{\mathscr{t}}

\def\secondMoment[#1]{\delta_{#1}}

\def\modelParametersAtIteration[#1]{\textbf{w}^{(#1)}}
\def\modelParametersAtIterationEle[#1][#2]{w^{(#1)}_{#2}}

\def\modelParameters{\textbf{w}}
\def\modelParametersEle[#1]{{w}_{#1}}
\def\modelParametersOptimal{{\textbf{w}^{*}}}

\def\localGradientSign[#1][#2]{\bar{\textbf{g}}_{#1}^{(#2)}}
\def\localGradient[#1][#2]{\tilde{\textbf{g}}_{#1}^{(#2)}}
\def\localGradientNoIndex[#1]{\tilde{\textbf{g}}_{#1}}
\def\localGradientAll[#1][#2]{{\textbf{g}}_{#1}^{(#2)}}

\def\localGradientElementQuantized[#1][#2]{{\bar{\tilde{g}}}_{#1}^{(#2)}}
\def\localGradientElement[#1][#2]{{\tilde{g}}_{#1}^{(#2)}}
\def\localGradientNoIndexElement[#1]{{\tilde{g}}_{#1}}
\def\gradientWeight[#1]{\omega_{#1}}

\def\encoderGradient[#1][#2][#3]{\psi_{#1}^{#2}(#3)}

\def\lossFunctionSample[#1]{f(#1)}
\def\lossFunctionLocal[#1][#2]{F_{#1}(#2)}
\def\lossFunctionGlobal[#1]{F(#1)}

\def\meanGradientEleEstimate[#1][#2]{{\hat{v}}^{(#1)}_{#2}}
\def\meanGradientEleOverQuantized[#1][#2]{{\bar{v}}^{(#1)}_{#2}}
\def\meanGradientEle[#1][#2]{{v}^{(#1)}_{#2}}
\def\meanGradientVector[#1]{\textbf{v}^{(#1)}}
\def\meanGradientVectorEstimate[#1]{\hat{\textbf{{v}}}^{(#1)}}

\def\globalGradient[#1]{{\textbf{{g}}}^{(#1)}}
\def\globalGradientElement[#1][#2]{{{g}}^{(#1)}_{#2}}

\def\globalGradientElementNoIndex[#1]{{g_{#1}}}
\def\bias[#1][#2]{{\textbf{{b}}}^{(#1)}_{#2}}

\def\metricForFirst[#1]{e_{#1}^{+}}
\def\metricForSecond[#1]{e_{#1}^{-}}

\def\nonnegativeConstantsEle[#1]{L_{#1}}

\def\varianceBoundEle[#1]{\sigma_{#1}}

\def\channelVector[#1]{\textbf{\textrm{h}}_{#1}^{(\indexCommunicationRound)}}

\def\receiveVector[#1]{\textbf{\textrm{r}}_{#1}^{(\indexCommunicationRound)}}
\def\receiveVectorEstimate[#1]{\tilde{\textbf{\textrm{{r}}}}_{#1}^{(\indexCommunicationRound)}}

\def\matrixForCut[#1]{\textbf{\textrm{C}}_{#1}}
\def\symbolVector[#1]{\textbf{\textrm{d}}_{#1}^{(\indexCommunicationRound)}}
\def\symbolVectorEstimate[#1]{\tilde{\textbf{\textrm{{d}}}}_{#1}^{(\indexCommunicationRound)}}
\def\receivedVector[#1]{\textbf{\textrm{r}}_{#1}^{(\indexCommunicationRound)}}
\def\noiseVector[#1]{\textbf{\textrm{n}}_{#1}^{(\indexCommunicationRound)}}
\def\noiseVectorOnSymbols[#1]{\tilde{\textbf{\textrm{n}}}_{#1}^{(\indexCommunicationRound)}}

\def\transmittedVector[#1]{\textbf{\textrm{t}}_{#1}^{(\indexCommunicationRound)}}
\def\channelVector[#1]{\textbf{\textrm{h}}_{#1}^{(\indexCommunicationRound)}}
\def\idftMatrix[#1]{\textbf{\textrm{F}}_{#1}^{\rm H}}
\def\dftMatrix[#1]{\textbf{\textrm{F}}_{#1}}
\def\transformPrecoder[#1]{\textbf{\textrm{T}}_{#1}}
\def\transformDecoder[#1]{\textbf{\textrm{T}}_{#1}^{\rm H}}
\def\dftPrecoder[#1]{\textbf{\textrm{D}}_{#1}}
\def\dftDecoder[#1]{\textbf{\textrm{D}}_{#1}^{\rm H}}

\def\channelImpulseResponse[#1]{\textbf{\textrm{h}}_{#1}^{(\indexCommunicationRound)}}
\def\channelMatrix[#1]{\textbf{\textrm{H}}_{#1}^{(\indexCommunicationRound)}}
\def\channelMatrixDiag[#1]{{\bf \Lambda}_{#1}^{(\indexCommunicationRound)}}

\def\numberOfAntennasAtES{R}

\def\syncError{T_{\rm sync}}

\def\numberOfParametersPerOFDM{M_{\rm par}}

\def\distanceED[#1]{r_{#1}}
\def\powerED[#1]{P_{#1}}

\def\indexArea{u}
\def\Nerror{N_{\text{err}}}

\newcommand\mydots{\hbox to 1em{.\hss.\hss.}}

\documentclass[10pt,conference,comsoc]{IEEEtran}
\usepackage{mathrsfs}
\usepackage{multicol}
\usepackage{lipsum}
\usepackage{mathtools}
\usepackage{amsthm}
\usepackage{acronym}
\usepackage[bookmarksopen=true]{hyperref}
\usepackage{stfloats}
\usepackage{amsfonts}
\usepackage{acronym}
\usepackage{cite}
\usepackage{multirow}
\usepackage{bm}
\usepackage{amsmath,amssymb}
\usepackage{graphicx}
\usepackage[table,xcdraw]{xcolor}
\usepackage[caption=false,font=footnotesize]{subfig}
\usepackage[geometry]{ifsym}
\usepackage{array}
\usepackage[utf8]{inputenc}
\usepackage[T1]{fontenc}
\let\norm\undefined % <-- "Undefine" \norm
\DeclarePairedDelimiter\norm{\lVert}{\rVert}

\usepackage{tikz}
\usepackage{lipsum}
\usetikzlibrary{calc,shadings,patterns}
\makeatletter
\tikzset{%
  remember picture with id/.style={%
    remember picture,
    overlay,
    save picture id=#1,
  },
  save picture id/.code={%
    \edef\pgf@temp{#1}%
    \immediate\write\pgfutil@auxout{%
      \noexpand\savepointas{\pgf@temp}{\pgfpictureid}}%
  },
  if picture id/.code args={#1#2#3}{%
    \@ifundefined{save@pt@#1}{%
      \pgfkeysalso{#3}%
    }{
      \pgfkeysalso{#2}%
    }
  }
}

\def\savepointas#1#2{%
  \expandafter\gdef\csname save@pt@#1\endcsname{#2}%
}

\def\tmk@labeldef#1,#2\@nil{%
  \def\tmk@label{#1}%
  \def\tmk@def{#2}%
}

\tikzdeclarecoordinatesystem{pic}{%
  \pgfutil@in@,{#1}%
  \ifpgfutil@in@%
    \tmk@labeldef#1\@nil
  \else
    \tmk@labeldef#1,(0pt,0pt)\@nil
  \fi
  \@ifundefined{save@pt@\tmk@label}{%
    \tikz@scan@one@point\pgfutil@firstofone\tmk@def
  }{%
  \pgfsys@getposition{\csname save@pt@\tmk@label\endcsname}\save@orig@pic%
  \pgfsys@getposition{\pgfpictureid}\save@this@pic%
  \pgf@process{\pgfpointorigin\save@this@pic}%
  \pgf@xa=\pgf@x
  \pgf@ya=\pgf@y
  \pgf@process{\pgfpointorigin\save@orig@pic}%
  \advance\pgf@x by -\pgf@xa
  \advance\pgf@y by -\pgf@ya
  }%
}

\makeatother
\newcounter{hatchNumber}
\DeclarePairedDelimiter\floor{\lfloor}{\rfloor}

\usepackage{cuted}
\makeatletter
\newif\ifAC@uppercase@first%
\def\Aclp#1{\AC@uppercase@firsttrue\aclp{#1}\AC@uppercase@firstfalse}%
\def\AC@aclp#1{%
	\ifcsname fn@#1@PL\endcsname%
	\ifAC@uppercase@first%
	\expandafter\expandafter\expandafter\MakeUppercase\csname fn@#1@PL\endcsname%
	\else%
	\csname fn@#1@PL\endcsname%
	\fi%
	\else%
	\AC@acl{#1}s%
	\fi%
}%
\def\Acp#1{\AC@uppercase@firsttrue\acp{#1}\AC@uppercase@firstfalse}%
\def\AC@acp#1{%
	\ifcsname fn@#1@PL\endcsname%
	\ifAC@uppercase@first%
	\expandafter\expandafter\expandafter\MakeUppercase\csname fn@#1@PL\endcsname%
	\else%
	\csname fn@#1@PL\endcsname%
	\fi%
	\else%
	\AC@ac{#1}s%
	\fi%
}%
\def\Acfp#1{\AC@uppercase@firsttrue\acfp{#1}\AC@uppercase@firstfalse}%
\def\AC@acfp#1{%
	\ifcsname fn@#1@PL\endcsname%
	\ifAC@uppercase@first%
	\expandafter\expandafter\expandafter\MakeUppercase\csname fn@#1@PL\endcsname%
	\else%
	\csname fn@#1@PL\endcsname%
	\fi%
	\else%
	\AC@acf{#1}s%
	\fi%
}%
\def\Acsp#1{\AC@uppercase@firsttrue\acsp{#1}\AC@uppercase@firstfalse}%
\def\AC@acsp#1{%
	\ifcsname fn@#1@PL\endcsname%
	\ifAC@uppercase@first%
	\expandafter\expandafter\expandafter\MakeUppercase\csname fn@#1@PL\endcsname%
	\else%
	\csname fn@#1@PL\endcsname%
	\fi%
	\else%
	\AC@acs{#1}s%
	\fi%
}%
\edef\AC@uppercase@write{\string\ifAC@uppercase@first\string\expandafter\string\MakeUppercase\string\fi\space}%
\def\AC@acrodef#1[#2]#3{%
	\@bsphack%
	\protected@write\@auxout{}{%
		\string\newacro{#1}[#2]{\AC@uppercase@write #3}%
	}\@esphack%
}%
\def\Acl#1{\AC@uppercase@firsttrue\acl{#1}\AC@uppercase@firstfalse}
\def\Acf#1{\AC@uppercase@firsttrue\acf{#1}\AC@uppercase@firstfalse}
\def\Ac#1{\AC@uppercase@firsttrue\ac{#1}\AC@uppercase@firstfalse}
\def\Acs#1{\AC@uppercase@firsttrue\acs{#1}\AC@uppercase@firstfalse}

\newtheorem{example}{\color{black} Example} 
\DeclareMathOperator{\sign}{sign}
\def\signNormal[#1]{\sign\left(#1\right)}

\DeclareMathOperator{\diag}{diag}
\def\diagOperator[#1]{\diag\left\{#1\right\}}

\acrodef{QSGD}{quantized \ac{SGD}}
\acrodef{SNR}{signal-to-noise ratio}
\acrodef{RMSE}{root-mean-square error}
\acrodef{OFDM}{orthogonal frequency division multiplexing}
\acrodef{DFT}{discrete Fourier transform}
\acrodef{PSK}{phase-shift keying}
\acrodef{QAM}{quadrature amplitude modulation}
\acrodef{QPSK}{quadrature phase-shift keying}
\acrodef{PMEPR}{peak-to-mean envelope power ratio}
\acrodef{BER}{bit-error ratio}
\acrodef{SNR}{signal-to-noise ratio}
\acrodef{PSD}{power spectral density}
\acrodef{SE}{spectral efficiency}
\acrodef{CP}{cyclic prefix}
\acrodef{AWGN}{additive white Gaussian noise}
\acrodef{CFR}{channel frequency response}
\acrodef{CIR}{channel impulse response}
\acrodef{MMSE}{minimum mean square error}
\acrodef{LMMSE}{linear minimum mean square error}
\acrodef{BPSK}{binary phase shift keying}
\acrodef{BLER}{block-error rate}
\acrodef{ML}{maximum likelihood}
\acrodef{PHY}{physical layer}
\acrodef{PA}{power amplifier}
\acrodef{IDFT}{inverse DFT}
\acrodef{DoF}{degrees-of-freedom}
\acrodef{IoT}{Internet-of-Things}
\acrodef{FDE}{frequency-domain equalization}
\acrodef{FSK}{frequency-shift keying}
\acrodef{FSK-MV}{\ac{FSK}-based \ac{MV}}
\acrodef{PPM}{pulse-position modulation}
\acrodef{PPM-MV}{\ac{PPM}-based \ac{MV}}
\acrodef{RF}{radio-frequency}
\acrodef{IM}{index modulation}
\acrodef{BS}{base station}
\acrodef{MF}{matched filter}

\acrodef{BAA}{broadband analog aggregation}
\acrodef{OBDA}{one-bit broadband digital aggregation}
\acrodef{FEEL}{federated edge learning}
\acrodef{FL}{federated learning}
\acrodef{ED}{edge device}
\acrodef{ES}{edge server}
\acrodef{UL}{uplink}
\acrodef{DL}{downlink}
\acrodef{OAC}{over-the-air computation}
\acrodef{TCI}{truncated-channel inversion}
\acrodef{MV}{majority vote}
\acrodef{CNN}{convolution neural network}
\acrodef{ReLU}{rectified-linear unit}
\acrodef{CSI}{channel state information}
\acrodef{PAPR}{peak-to-average power ratio}
\acrodef{iid}[IID]{independent and identically distributed}
\acrodef{5G}{Fifth Generation}
\acrodef{4G}{Fourth Generation}
\acrodef{NR}{New Radio}
\acrodef{LTE}{Long Term Evolution}
\acrodef{RACH}{random-access channel}
\acrodef{DNN}{deep nueral network}
\acrodef{SGD}{stochastic gradient descent}
\acrodef{signSGD}{sign stochastic gradient descent}
\acrodef{5G}{Fifth Generation}
\acrodef{4G}{Fourth Generation}
\acrodef{NR}{New Radio}
\acrodef{LTE}{Long Term Evolution}
\acrodef{PRACH}{physical random access channel}
\acrodef{PUCCH}{physical uplink control channel}
\acrodef{OFDMA}{orthogonal frequency division multiple access}

\acrodef{MRC}{maximum-ratio combining}

\acrodef{MD}{multi-digit}
\acrodef{BMSE}{Bayesian \ac{MSE}}
\acrodef{MSE}{mean squared error}
\acrodef{NMSE}{normalized \ac{MSE}}
\acrodef{AAM}{adaptive absolute maximum}

\def\BibTeX{{\rm B\kern-.05em{\sc i\kern-.025em b}\kern-.08em
    T\kern-.1667em\lower.7ex\hbox{E}\kern-.125emX}}

\begin{document}

\title{Over-the-Air Computation over Balanced Numerals} 

\author{\IEEEauthorblockN{Alphan \c{S}ahin\IEEEauthorrefmark{1} and Rui Yang\IEEEauthorrefmark{2}}
	\IEEEauthorblockA{\IEEEauthorrefmark{1}University of South Carolina, Columbia, SC, USA, 
		Email: asahin@mailbox.sc.edu}
	\IEEEauthorblockA{\IEEEauthorrefmark{2}InterDigital, New York, NY, USA, Email: rui.yang@interdigital.com}}

%\author{
%	\IEEEauthorblockN{Alphan \c{S}ahin and Rui Yang}
%	\IEEEauthorblockA{ and uadr.\\
%		Emails: }
%}

\maketitle

\begin{abstract}
In this study, a digital \ac{OAC} scheme for achieving continuous-valued gradient aggregation is proposed. It is shown that the average of a set of real-valued parameters can be calculated approximately by using the average of the corresponding numerals, where the numerals are obtained based on a balanced number system.  By using this property, the proposed scheme encodes the local gradients into a set of numerals. It then determines the positions of the activated  \ac{OFDM} subcarriers by using the values of the numerals. To eliminate the need for a precise sample-level time synchronization, channel estimation overhead, and power instabilities due to the channel inversion, the proposed scheme also uses a non-coherent receiver at the \ac{ES} and does not utilize a pre-equalization at the \acp{ED}. Finally, the theoretical \ac{MSE} performance of the proposed scheme is derived and its performance for \ac{FEEL} is demonstrated.
\end{abstract}
\acresetall
\section{Introduction}
\Ac{OAC} aims to compute nomographic functions by exploiting the signal-superposition property of wireless multiple-access channels \cite{Nazer_2007, Gastpar_2003}. \ac{OAC}, which initially considered for wireless sensor networks \cite{Goldenbaum_2013}, has recently gained increasing  attention for applications such as distributed learning  or wireless control systems \cite{Wanchun_2020,chen2021distributed,Park_2021}. For example, \ac{FEEL}, one of the promising distributed edge learning frameworks, aims to implement \ac{FL} \cite{pmlr-v54-mcmahan17a} over a wireless network. With \ac{FEEL}, the task of model training  is distributed across multiple \acp{ED} and the data uploading is avoided to promote user-privacy \cite{Mingzhe_2021,chen2021distributed}. Instead of data samples, \acp{ED} share a large number of local stochastic gradients (or local model parameters) with an \ac{ES} for aggregation, e.g., averaging. However, typical orthogonal user multiplexing methods such as \ac{OFDMA} can be wasteful in this scenario since the \ac{ES} may  not be interested in the local information of the \acp{ED} but only in a function of them.% Similarly, a control system that requires an input that is a function of many \ac{IoT} devices' readings can suffer from high latency since the available spectrum for these networks is often limited and the \ac{OAC} can address the latency issues by calculating the functions, e.g., difference equations \cite{Park_2021}, over the air.

Although \ac{OAC} is a promising concept to address the latency issues for \ac{FEEL}, it is challenging to realize a reliable \ac{OAC} scheme due to the  detrimental impact of wireless channels on the \ac{OAC}. To address this issue, a majority of the state-of-the-art  \ac{OAC} methods relying on pre-equalization techniques\cite{Guangxu_2020, sery2020overtheair, Amiri_2020,Guangxu_2021, Zhang_2021}. However, a pre-equalizer can impose stringent requirements on the underlying mechanisms such as time-frequency synchronization, channel estimation, and channel prediction, which can be challenging to satisfy under the non-stationary  channel conditions \cite{Haejoon_2021}.
%However, to implement  channel inversion techniques  in a practical wireless network, it is a hard problem due to non-stationary behavior of the channel and 
%Also, the channel coefficients quickly change due to the mobility in the environment, the phase drifts caused by the carrier frequency offset, and time-synchronization errors at both \acp{ED} and the \ac{ES}. In addition, for a large number of parameters, the pre-equalization coefficients may need to be updated periodically within the transmission.
%, which is not trivial to implement. 
%Also, the typical equalization or phase correction methods used at the receiver for traditional multiple-access schemes, e.g., \ac{OFDMA}, cannot be directly employed for \ac{OAC}  to compensate the channel distortion or imperfect synchronization as the impact of distortions on the received symbols after  the superposition often do not satisfy the distributive law of mathematical operations. A
Another issue is that most of the \ac{OAC} schemes use analog modulation schemes to achieve a continuous-valued computation. However, analog modulations are more susceptible to noise as compared to digital schemes. Although there are  digital aggregation methods, e.g., \ac{OBDA} \cite{Guangxu_2021}, \ac{FSK-MV} \cite{sahinCommnet_2021}, and \ac{PPM-MV} \cite{sahinWCNC_2022}, by relying on specific training approaches, i.e., distributed training by the \ac{MV} with \ac{signSGD} \cite{Bernstein_2018}, these schemes do not allow one to compute a continuous-valued function.

% In addition,  channel  under the fading channel can cause instantaneous power fluctuations.

In this paper, to address the aforementioned issues, we introduce an \ac{OAC} scheme for \ac{FEEL}, where the local gradients are encoded into a set of numerals  based on a balanced number system (also called balanced numeral system  or signed-digit representation)    \cite{koren2018computer} to achieve a continuous-valued computation over  a digital scheme.\footnote{To avoid confusion, we use the terms of "numeral" and "balanced" for "digit" and "signed-digit", respectively, since the term of "digit" may specifically imply the ten symbols of the common base 10 numeral system.} The proposed method does not rely on pre-equalization and the availability of \ac{CSI} at the \acp{ED} and the \ac{ES}, which relaxes the  synchronization requirements at the \acp{ED} and the \ac{ES}. We demonstrate the efficacy of the proposed \ac{OAC} scheme for both homogeneous and heterogeneous data distribution scenarios.
% needed for the methods relying on a channel inversion technique 

{\em Notation:} The sets of complex numbers, real numbers, integers, and integers modulo $H$ are denoted by $\complexNumbers$,  $\realNumbers$, $\integers$, and $\integers_H$ respectively.
The $N$-dimensional all zero vector and the $N\times N$ identity matrix are  $\zeroVector[{N}]$ and $\identityMatrix[{N}]$, respectively. 
The function $\indicatorFunction[\cdot]$ results in $1$ if its argument holds, otherwise it is $0$. 
$\expectationOperator[\cdot][]$ is the expectation of its  argument. 
 %The operation $\diagOperator[\textbf{a}]$  returns a square diagonal matrix with the elements of vector \textbf{a} on the main diagonal.
$ \nabla 
 \lossFunctionSample[{\modelParameters}]$ denotes the gradient of the function $f$, i.e. $\nabla f$, at the point $\modelParameters$. 
The zero-mean circularly symmetric multivariate  complex Gaussian distribution with the covariance matrix ${\textbf{\textrm{C}}_{\numberOfActiveSubcarriers}}$ of an $\numberOfActiveSubcarriers$-dimensional random vector $\vectorX\in\complexNumbers^{\numberOfActiveSubcarriers\times1}$ is denoted by
$\vectorX\sim\complexGaussian[\zeroVector[\numberOfActiveSubcarriers]][{\textbf{\textrm{C}}_{\numberOfActiveSubcarriers}}]$. The gamma distribution with the shape parameter $n$ and the rate  $\lambda$ is  $\gammaDist[n][\lambda]$.
 The $\ell_2$-norm of the vector $\vectorX$ is $\norm{\vectorX}_2$.

\section{System Model}
\label{sec:system}

We consider a  network with $\numberOfEdgeDevices$ \acp{ED} that are connected to an \ac{ES} wirelessly, where each \ac{ED} and the \ac{ES}  are equipped with a single antenna and $\numberOfAntennasAtES$ antennas, respectively. We assume that the large-scale impact of the wireless channel is compensated with a power control mechanism to maintain the average received signal power of each \ac{ED} at the ES identical. %As our focus is to provide insight into the proposed \ac{OAC} scheme, we do not consider the impact of multiple cells on the proposed \ac{OAC} (e.g., inter-cell interference).
 %\subsection{Signal Model}
For the signal model, we assume that the \acp{ED} access the wireless channel  on the same time-frequency resources {\em simultaneously} with $\numberOfOFDMSymbols$ \ac{OFDM} symbols consisting of $\numberOfActiveSubcarriers$ active subcarriers  for \ac{OAC}.  %We assume the transmissions from the \acp{ED} are synchronized in both time and frequency and arrive at the \ac{ES} within the \ac{CP} duration. 
 Assuming that the \ac{CP} duration is larger than the sum of maximum time-synchronization error and maximum-excess delay of the channel,
 %the symbol-level synchronization  among the \acp{ED} signals at the \ac{ES} is achieved via a \ac{CP} duration larger the maximum-excess delay of the multipath channel and providing enough overhead for synchronization.  
we express the received symbol on  the $\indexSubcarrier$th subcarrier of the $\indexOFDMSymbol$th \ac{OFDM} symbol at the \ac{ES}  for the $\indexCommunicationRound$th communication  round of the training as
 \begin{align}
 	\receivedSymbolAtSubcarrier[{\indexSubcarrier,\indexOFDMSymbol}] = \sum_{\indexED=0}^{\numberOfEdgeDevices-1} \channelAtSubcarrier[\indexED,\indexSubcarrier,\indexOFDMSymbol]\transmittedSymbolAtSubcarrier[\indexED,\indexSubcarrier,\indexOFDMSymbol]+\noiseAtSubcarrier[\indexSubcarrier,\indexOFDMSymbol]~,
 	\label{eq:symbolOnSubcarrier}
 \end{align}
 where  $\channelAtSubcarrier[\indexED,\indexSubcarrier,\indexOFDMSymbol]\sim\complexGaussian[\zeroVector[\numberOfAntennasAtES]][{\identityMatrix[\numberOfAntennasAtES]}]$ is a vector that consists of the channel coefficients between $\numberOfAntennasAtES$ antennas at the \ac{ES} and the $\indexED$th \ac{ED}, $\transmittedSymbolAtSubcarrier[\indexED,\indexSubcarrier,\indexOFDMSymbol]\in\complexNumbers$ is the transmitted modulation symbol from the $\indexED$th \ac{ED}, and ${\noiseAtSubcarrier[\indexSubcarrier,\indexOFDMSymbol]}\sim\complexGaussian[\zeroVector[\numberOfAntennasAtES]][{\noiseVariance\identityMatrix[\numberOfAntennasAtES]}]$ is the \ac{AWGN}, where  $\noiseVariance$ is the noise variance for $\indexSubcarrier\in\integers_\numberOfActiveSubcarriers$ and $\indexOFDMSymbol\in\integers_\numberOfOFDMSymbols$.   We denote the \ac{SNR} of an \ac{ED} at the \ac{ES} receiver as $1/\noiseVariance$. 
 
 In practice, the synchronization point where the \ac{DFT} starts to be applied to the received signal for demodulation at the \ac{ES} and the time synchronization across the \acp{ED} may not be precise. To model former impairment, we assume that synchronization point can deviate by $\Nerror$ samples within the \ac{CP} window. For the latter impairment, the time of arrivals of the \acp{ED}' signals at the \ac{ES} location are sampled from a uniform distribution between $0$ and $\syncError$~seconds, where $\syncError$ is equal to the reciprocal to the signal bandwidth. Note the coarse time-synchronization can be maintained with the state-of-the-art protocols used in cellular systems. We introduce additional phase rotations to $\channelAtSubcarrier[\indexED,\indexSubcarrier,\indexOFDMSymbol]$ to capture the impact of the time-synchronization errors on $\receivedSymbolAtSubcarrier[{\indexSubcarrier,\indexOFDMSymbol}]$. %We assume that the frequency synchronization is handled before the transmissions. 
 
 \subsection{Balanced Number Systems}
 \label{subsec:quanBalan}
 \def\base{\beta}
 \def\indexDigit{i}
 \def\baseBT{\text{b3}}
 \def\numberOfDigits{D}
 \def\normalizationDigit{\xi}
 \def\valueMaximum{v_{\rm max}}
 \def\representationInBase[#1][#2]{\left(#2\right)}
 \def\digitAveraged[#1][#2]{\mu_{{#1},{#2}}^{(\indexCommunicationRound)}}
 \def\digitAveragedEst[#1][#2]{\hat{\mu}_{{#1},{#2}}^{(\indexCommunicationRound)}}
 \def\digitGeneral[#1]{x_{#1}}
 \def\digit[#1][#2]{d_{{#1},{#2}}^{(\indexCommunicationRound)}}
 \def\digitStandart[#1]{b_{#1}}
 
 \def\symbolSetWithoutZero{\mathbb{S}_\base^0}
 \def\symbolWithoutZero[#1]{b_{#1}}
 \def\resourceSet[#1]{\mathbb{T}_{#1}}
 
 \def\symbolSet[#1]{\mathbb{S}_#1}
 \def\symbol[#1]{a_{#1}}
 \def\indexSymbol{j}
 \def\decoder{f_{\text{dec},\base}}
 \def\encoder{f_{\text{enc},\base}}
 \def\aRealValue{v}
 \def\aQuantizedValue{\bar{v}}
 \def\encoderSB{f_{\text{enc}}^{\text{bin}}}
 \def\encoderBT{f_{\text{enc}}^{\text{ter}}}
 \def\stepSize{\Delta}
 \def\functionBijection[#1]{f_{\rm bal}(#1)}

 Let $\encoder$ be a function that maps $\aRealValue\in\realNumbers$ to a sequence  of $\numberOfDigits$ elements (i.e., numerals) in $\{\representationInBase[\base][{\digitGeneral[\numberOfDigits-1],\mydots,\digitGeneral[1],\digitGeneral[0]}]|\digitGeneral[\indexDigit]\in\symbolSet[\base],\base>1, \indexDigit\in\integers_\numberOfDigits\}$ as 
 \begin{align}
 	\representationInBase[\base][{\digitGeneral[\numberOfDigits-1],\mydots,\digitGeneral[1],\digitGeneral[0]}]=\encoder(\aRealValue)~,
 \end{align}
 where $\base$ is the base (also called scale \cite{hardy08}), $\digitGeneral[\indexDigit]$ is referred to as a numeral at the $\indexDigit$th position,  and $\symbolSet[\base]$ is the symbol set with base $\base$. In this study, we  consider a balanced number system for expressing $\encoder$ and assume that $\base$ is an odd positive integer. The numerals are obtained as follows: For a given $\aRealValue$ for $|\aRealValue|\le\valueMaximum$, the encoder $\encoder(\aRealValue)$  first  computes the base-$\base$ representation of the rounded, biased, and normalized $\aRealValue$ as
 \begin{align}
 	\floor*{\frac{\normalizationDigit}{\valueMaximum}\aRealValue+\normalizationDigit+\frac{1}{2}}\triangleq\sum_{\indexDigit=0}^{\numberOfDigits-1}\digitStandart[\indexDigit]\base^\indexDigit~,
 	\label{eq:unternaryConversion}
 \end{align}
 for $\normalizationDigit\triangleq{(\base^{\numberOfDigits}-1)}/{2}$ and $\digitStandart[\indexDigit]\in\integers_\base$. Afterwards, it calculates   $\digitGeneral[\indexDigit]$ as $
 \digitGeneral[\indexDigit]\triangleq\digitStandart[\indexDigit]-(\base-1)/2$, $\forall\indexDigit\in\integers_\numberOfDigits$. Hence, $\symbolSet[\base]$ can be defined as 
 \begin{align}
 	\symbolSet[\base]
 	\triangleq\{\symbol[{\indexSymbol}]|\symbol[{\indexSymbol}]=\functionBijection[\indexSymbol], \indexSymbol\in\integers_\base\}~,
 	\label{eq:setDef}
 \end{align} 
where $\functionBijection[\indexSymbol]$ is given by
 \begin{align}
 	\functionBijection[\indexSymbol]\triangleq\begin{cases}
 		-(\indexSymbol+1)/2, & \text{odd } \indexSymbol, \indexSymbol<\base-1\\
 		(\indexSymbol+2)/2, & \text{even } \indexSymbol, \indexSymbol<\base-1\\
 		0, & \indexSymbol=\base-1
 	\end{cases}~.
 	\label{eq:bal}
 \end{align}
 Based on \eqref{eq:bal}, $\symbol[{\base-1}]$ is a zero-valued symbol. The example symbol sets for $\base=5$ and $\base=7$ can obtained as $\symbolSet[5]=\{-1,1,-2,2,0\}$ and $\symbolSet[7]=\{-1,1,-2,2,-3,3,0\}$  , respectively.  For a balanced number system, there is no dedicated symbol for the sign of $\aRealValue$ as $\symbolSet[\base]$ contains negative-valued symbols. 
 
 \begin{example}
 	\rm
 	Assume that $\base=5$, $\numberOfDigits=3$, and $\valueMaximum=1$ and we want to calculate $\encoder(0.28)$ and $\encoder(-0.86)$. By the definition, $\normalizationDigit=(5^2-1)/2=62$.
 	The base 5 representations of the decimal $\floor{62\times0.28+62+1/2}=79$ and the decimal $\floor{62\times-0.86+62+1/2}=9$  are $({\digitStandart[2]\digitStandart[1]\digitStandart[0]})_5=(304)_5$ and $({\digitStandart[2]\digitStandart[1]\digitStandart[0]})_5=(014)_5$, respectively. Since $
 	\digitGeneral[\indexDigit]\triangleq\digitStandart[\indexDigit]-(\base-1)/2$, we obtain $\encoder(0.28)=\representationInBase[5][1,-2,2]$, and $\encoder(-0.86)=\representationInBase[5][-2,-1,2]$.
 	\label{ex:enc}
 	\vspace{-2mm}
 \end{example}

 The corresponding decoder $\decoder$ that maps the sequence $\representationInBase[\base][{\digitGeneral[\numberOfDigits-1],\mydots,\digitGeneral[1],\digitGeneral[0]}]$  to $\aQuantizedValue\in\realNumbers$ can be expressed as
 \begin{align}
 	\aQuantizedValue=
 	\decoder\representationInBase[\base][{\digitGeneral[\numberOfDigits-1],\mydots,\digitGeneral[1],\digitGeneral[0]}]\triangleq\frac{\valueMaximum}{\normalizationDigit}\sum_{\indexDigit=0}^{\numberOfDigits-1}\digitGeneral[\indexDigit]\base^{\indexDigit}~.
 	\label{eq:decoderFcn}
 \end{align}

 Note that $\aQuantizedValue=\decoder\encoder(\aRealValue)$ forms a mid-tread uniform quantization, i.e., zero is one of the re-construction levels. The quantization step size can also be calculated as $\stepSize=2\valueMaximum/(\base^\numberOfDigits-1)$ and the quantization error, i.e., $|{\aRealValue-\aQuantizedValue}|$, decreases with increasing $\numberOfDigits$ for $|\aRealValue|\le\valueMaximum$. 
 %The mid-riser uniform quantization can also be used for the proposed \ac{OAC} approach.
 
 \begin{example}
 	\rm
 	Consider the parameters given in Example~\ref{ex:enc}. Hence, we obtain $\decoder\encoder(0.28)=\decoder\representationInBase[5][1,-2,2]\approxeq0.2742$, and $\decoder\encoder(-0.86)=\decoder\representationInBase[5][-2,-1,2]\approxeq-0.8548$ based on \eqref{eq:decoderFcn}. The step size can also be calculated as $\stepSize=2/(5^3-1)\approxeq0.016$. 
 	\label{ex:dec}
	 	\vspace{-2mm}
 \end{example}

 \subsection{Learning Model}
 \label{subsec:feel}
 Let $\dataset[\indexED]$ denote the local data set containing the labeled data samples $(\sampleData[\indexSampleData], \sampleLabel[\indexSampleData] )$ at the $\indexED$th \ac{ED}, $\forall\indexED\in\integers_\numberOfEdgeDevices$, where $\sampleData[\indexSampleData]$ is the $\indexSampleData$th data sample with its ground truth  label $\sampleLabel[\indexSampleData]$. Suppose that all \acp{ED} upload their data sets to the \ac{ES}. The centralized learning problem can then  be expressed as 
 \begin{align}
 	\modelParametersOptimal=\arg\min_{\modelParameters\in\realNumbers^{\numberOfModelParameters}} \lossFunctionGlobal[\modelParameters]=\arg\min_{\modelParameters\in\realNumbers^{\numberOfModelParameters}} \frac{1}{|\completeData|}\sum_{\forall(\sampleData[\indexSampleData],\sampleLabel[\indexSampleData])\in\completeData} \lossFunctionSample[{\modelParameters;\sampleData[\indexSampleData],\sampleLabel[\indexSampleData]}]~,
 	\nonumber
 \end{align}
 where $\lossFunctionGlobal[\modelParameters]$ is the loss function,  $\completeData=\dataset[0]\cup\dataset[1]\cup\cdots\cup\dataset[\numberOfEdgeDevices-1]$ is complete data set, and  $\lossFunctionSample[{\modelParameters;\sampleData[\indexSampleData],\sampleLabel[\indexSampleData]}]$ is the sample loss function for the parameters $\modelParameters=[\modelParametersEle[0],\mydots,\modelParametersEle[\numberOfModelParameters-1]]^{\rm T}\in\realNumbers^{\numberOfModelParameters}$, and $\numberOfModelParameters$ is the number of parameters. With (full-batch) gradient descent, a local optimum point can be  obtained as
 \begin{align}
 	\modelParametersAtIteration[\indexCommunicationRound+1] = \modelParametersAtIteration[\indexCommunicationRound] - \learningRate  \globalGradient[\indexCommunicationRound]~,
 	\label{eq:gd}
 \end{align}
 where $\learningRate$ is the learning rate and the gradient vector $\globalGradient[\indexCommunicationRound]=[\globalGradientElement[\indexCommunicationRound][0],\mydots,\globalGradientElement[\indexCommunicationRound][\numberOfModelParameters-1]]^{\rm T}\in\realNumbers^{\numberOfModelParameters}$ can be expressed as 
 \begin{align}
 	\globalGradient[\indexCommunicationRound] =  \nabla \lossFunctionGlobal[{\modelParametersAtIteration[\indexCommunicationRound]}]
 	= \frac{1}{|\completeData|}\sum_{\forall(\sampleData[\indexSampleData],\sampleLabel[\indexSampleData])\in\completeData} \nabla 
 	\lossFunctionSample[{\modelParametersAtIteration[\indexCommunicationRound];\sampleData[\indexSampleData],\sampleLabel[\indexSampleData]}]
 	~.
 	\label{eq:GlobalGradient}
 \end{align}
 %where the $\indexGradient$th element of the vector $\globalGradient[\indexCommunicationRound]$ is the gradient  of $\lossFunctionGlobal[{\modelParametersAtIteration[\indexCommunicationRound]}]$ with respect to $\modelParametersAtIterationEle[\indexCommunicationRound][\indexGradient]$. 
The equation \eqref{eq:gd} can be re-written as
  \begin{align}
 		\modelParametersAtIteration[\indexCommunicationRound+1] &= \modelParametersAtIteration[\indexCommunicationRound] - \learningRate \sum_{\indexED=0}^{\numberOfEdgeDevices-1}{\frac{|\dataset[\indexED]|}{|\completeData|}}\underbrace{\frac{1}{|\dataset[\indexED]|}\sum_{\forall(\sampleData[\indexSampleData],\sampleLabel[\indexSampleData]) \in\dataset[\indexED]} \nabla 
 			\lossFunctionSample[{\modelParametersAtIteration[\indexCommunicationRound];\sampleData[\indexSampleData],\sampleLabel[\indexSampleData]}]}_{\triangleq\localGradientAll[\indexED][\indexCommunicationRound]}~,\nonumber
 		%\nonumber\\
% 		&= \sum_{\indexED=0}^{\numberOfEdgeDevices-1}{\frac{|\dataset[\indexED]|}{|\completeData|}}\left(\modelParametersAtIteration[\indexCommunicationRound] - \learningRate \localGradientAll[\indexED][\indexCommunicationRound]\right) 	\nonumber
 	%\label{eq:gradientAgg}
 \end{align}
 where $\localGradientAll[\indexED][\indexCommunicationRound]\in\realNumbers^{\numberOfModelParameters}$ denotes the local gradient vector  at the $\indexED$th \ac{ED}. Therefore, \eqref{eq:gd} can still be realized by communicating the local gradients or locally updated model parameters between the \acp{ED} and the \ac{ES}, rather than moving the local data sets from the \acp{ED} to the \ac{ES}, which is beneficial for promoting data privacy \cite{Mingzhe_2021,chen2021distributed}. It also shows the underlying principle of the plain \ac{FL} based on gradient or parameter aggregations \cite{pmlr-v54-mcmahan17a}. 
 
 \ac{FEEL} aims to realize \ac{FL} over a wireless network. In this study, we consider \ac{FL} based on \ac{SGD}, known as FedSGD  \cite{pmlr-v54-mcmahan17a}, over a wireless network:  The $\indexED$th \ac{ED} calculates an estimate of the local gradients, denoted by $\localGradient[\indexED][\indexCommunicationRound]=[\localGradientElement[\indexED,0][\indexCommunicationRound],\mydots,\localGradientElement[\indexED,\numberOfModelParameters-1][\indexCommunicationRound]]^{\rm T}\in\realNumbers^{\numberOfModelParameters}$, as 
 \begin{align}
 	\localGradient[\indexED][\indexCommunicationRound] =  \nabla  \lossFunctionLocal[\indexED][{\modelParametersAtIteration[\indexCommunicationRound]}] 
 	= \frac{1}{\batchSize} \sum_{\forall(\sampleData[\indexSampleData],\sampleLabel[\indexSampleData])\in\datasetBatch[\indexED]} \nabla 
 	\lossFunctionSample[{\modelParametersAtIteration[\indexCommunicationRound];\sampleData[\indexSampleData],\sampleLabel[\indexSampleData]}]
 	~,
 	\label{eq:LocalGradientEstimate}
 \end{align}
 where $\datasetBatch[\indexED]\subset\dataset[\indexED]$ is the data batch obtained from the local data set and $\batchSize=|\datasetBatch[\indexED]|$ as the batch size. The \acp{ED} transmit the local gradient estimates to the \ac{ES}. Assuming identical data set sizes across the \acp{ED},  the \ac{ES} calculates the average stochastic gradient vector $\meanGradientVector[\indexCommunicationRound]\triangleq[\meanGradientEle[\indexCommunicationRound][0],\mydots,\meanGradientEle[\indexCommunicationRound][\numberOfModelParameters-1]]^{\rm T}=\frac{1}{\numberOfEdgeDevices} \sum_{\indexED=0}^{\numberOfEdgeDevices-1}\localGradient[\indexED][\indexCommunicationRound]$ and broadcasts it to the \acp{ED}. Finally, the model parameters at the \acp{ED} are updated as $ 	\modelParametersAtIteration[\indexCommunicationRound+1] = \modelParametersAtIteration[\indexCommunicationRound] - \learningRate\meanGradientVector[\indexCommunicationRound]$.

With a traditional orthogonal user multiplexing, the per-round communication latency for \ac{FEEL} linearly increases with the number of \acp{ED} \cite{liu2021training}.
With the motivation of eliminating per-round communication latency, the main objective of this work is to calculate an estimate of $\meanGradientVector[\indexCommunicationRound]$, denoted by $\meanGradientVectorEstimate[\indexCommunicationRound]\triangleq[\meanGradientEleEstimate[\indexCommunicationRound][0],\mydots,\meanGradientEleEstimate[\indexCommunicationRound][\numberOfModelParameters-1]]^{\rm T}$, through a digital \ac{OAC} scheme robust against fading channel.

\section{Proposed OAC Scheme}
\label{subsec:key}
Based on the discussions given in Section~\ref{subsec:feel}, consider the $\indexGradient$th gradient at the $\indexED$th \ac{ED} for the $\indexCommunicationRound$th communication round of the \ac{FEEL}, i.e., $\localGradientElement[\indexED,\indexGradient][\indexCommunicationRound]$. For a given $\base$, suppose that  $\localGradientElement[\indexED,\indexGradient][\indexCommunicationRound]$ is encoded into the sequence of length $\numberOfDigits$ denoted by
\begin{align}
\representationInBase[\base][{\digit[\indexED,\indexGradient][\numberOfDigits-1],\mydots,\digit[\indexED,\indexGradient][1],\digit[\indexED,\indexGradient][0]}]=\encoder(\localGradientElement[\indexED,\indexGradient][\indexCommunicationRound])~,
\label{eq:encodedSequence}
\end{align}
for $\digit[\indexED,\indexGradient][\indexDigit]\in\symbolSet[\base]$.
%Assume that the \ac{ES} needs to calculate the averaged stochastic gradient for the $\indexGradient$th parameter across $\numberOfEdgeDevices$ \acp{ED}. 
By using definition of $\decoder$ in \eqref{eq:decoderFcn}, the $\indexGradient$th average stochastic gradient, i.e., $\meanGradientEle[\indexCommunicationRound][\indexGradient]=
\frac{1}{\numberOfEdgeDevices}\sum_{\indexED=0}^{\numberOfEdgeDevices-1} \localGradientElement[\indexED,\indexGradient][\indexCommunicationRound]$,  can be obtained approximately as
\begin{align}
	\meanGradientEle[\indexCommunicationRound][\indexGradient]\approxeq\meanGradientEleOverQuantized[\indexCommunicationRound][\indexGradient]\triangleq\frac{1}{\numberOfEdgeDevices}\sum_{\indexED=0}^{\numberOfEdgeDevices-1}& \localGradientElementQuantized[\indexED,\indexGradient][\indexCommunicationRound]=\frac{\valueMaximum}{\normalizationDigit}\sum_{\indexDigit=0}^{\numberOfDigits-1} \underbrace{\frac{1}{\numberOfEdgeDevices}\sum_{\indexED=0}^{\numberOfEdgeDevices-1}\digit[\indexED,\indexGradient][\indexDigit]}_{\triangleq\digitAveraged[\indexGradient][\indexDigit]}\base^{\indexDigit}\nonumber
\\
&=\decoder\representationInBase[\base][{\digitAveraged[\indexGradient][\numberOfDigits-1],\mydots,\digitAveraged[\indexGradient][1],\digitAveraged[\indexGradient][0]}]~,
\label{eq:basicAveraging}
\end{align}
where $\localGradientElementQuantized[\indexED,\indexGradient][\indexCommunicationRound]$ is the quantized gradient, i.e.,
$\localGradientElementQuantized[\indexED,\indexGradient][\indexCommunicationRound] = \decoder\encoder(\localGradientElement[\indexED,\indexGradient][\indexCommunicationRound])$. Equation \eqref{eq:basicAveraging} implies that $\meanGradientEle[\indexCommunicationRound][\indexGradient]$ can be calculated approximately by evaluating the function $\decoder$ with the values that are calculated by averaging the numerals across $\numberOfEdgeDevices$ \acp{ED} in {\em real} domain, i.e., $\{\digitAveraged[\indexGradient][\indexDigit]|\indexDigit\in\integers_\numberOfDigits\}$. By evaluating $\digitAveraged[\indexGradient][\indexDigit]$ further, it can also be shown that
\begin{align}
\digitAveraged[\indexGradient][\indexDigit] = \frac{1}{\numberOfEdgeDevices}\sum_{\indexED=0}^{\numberOfEdgeDevices-1}\digit[\indexED,\indexGradient][\indexDigit] = \frac{1}{\numberOfEdgeDevices}\sum_{\indexSymbol=0}^{\base-1} \symbol[\indexSymbol]\numberOFEDsForOptionGeneral[\indexSymbol]~,
\label{eq:digitAveraging}
\end{align}
where $\numberOFEDsForOptionGeneral[\indexSymbol]$ denotes the number of \acp{ED} with the symbol $\symbol[\indexSymbol]$ for the $\indexDigit$th numeral in \eqref{eq:encodedSequence} and the $\indexGradient$th gradient. Note that the identity in \eqref{eq:digitAveraging} is due to the definition of expectation for discrete outcomes as given for a probability mass function.

\begin{example}
	\rm 
Assume that $\numberOfEdgeDevices=2$, $\localGradientElement[0,\indexGradient][\indexCommunicationRound]=0.28$, and $\localGradientElement[1,\indexGradient][\indexCommunicationRound]=-0.86$. The average of the gradients can be calculated as $\meanGradientEle[\indexCommunicationRound][\indexGradient]=(\localGradientElement[0,\indexGradient][\indexCommunicationRound]+\localGradientElement[1,\indexGradient][\indexCommunicationRound])/2=-0.29$. Now, consider the encoder parameters given in Example~\ref{ex:enc}. We obtain $\encoder(0.28)=\representationInBase[5][1,-2,2]$, and $\encoder(-0.86)=\representationInBase[5][-2,-1,2]$. Therefore, the average of the numerals can be calculated as $(\digitAveraged[\indexGradient][2],\digitAveraged[\indexGradient][1],\digitAveraged[\indexGradient][0])=(1-2,-2-1,2+2)/2=(-1/2,-3/2,2)$. Also, notice that $(\digitAveraged[\indexGradient][2],\digitAveraged[\indexGradient][1],\digitAveraged[\indexGradient][0])$ can be calculated by using the number of \acp{ED} that votes for each element of $\{-1,1,-2,2,0\}$. For instance, $\digitAveraged[\indexGradient][0]$ can be calculated via the last expression in \eqref{eq:digitAveraging} for
$(\numberOFEDsForOptionGeneral[0],\numberOFEDsForOptionGeneral[1],\numberOFEDsForOptionGeneral[2],\numberOFEDsForOptionGeneral[3],\numberOFEDsForOptionGeneral[4])=(0,0,0,2,0)$ where the corresponding symbols are $(\symbol[0],\symbol[1],\symbol[2],\symbol[3],\symbol[4])=(-1,1,-2,2,0)$ for $\base=5$. Finally, by evaluating $\meanGradientEleOverQuantized[\indexCommunicationRound][\indexGradient]=\decoder(-1/2,-3/2,2)$, we obtain $\meanGradientEleOverQuantized[\indexCommunicationRound][\indexGradient]\approxeq-0.2903$. Note that $\meanGradientEleOverQuantized[\indexCommunicationRound][\indexGradient]$ is also equal to the average of the quantized gradients, i.e., $\decoder\encoder(0.28)\approxeq0.2742$ and $\decoder\encoder(-0.86)\approxeq-0.8548$, as provided in Example~\ref{ex:dec}.  
\label{ex:sum}
\end{example}

The main equations used in this work are the encoding in \eqref{eq:encodedSequence}, the expansion in \eqref{eq:basicAveraging}, and the identity in \eqref{eq:digitAveraging}, which enable one to calculate an estimate of $\meanGradientEle[\indexCommunicationRound][\indexGradient]$ by averaging numerals, rather than the actual values of the gradients.

\subsection{Edge Device - Transmitter}
%The proposed scheme relies on a low-complexity \ac{FSK} over \ac{OFDM}: 
At the $\indexCommunicationRound$th communication round of the \ac{FEEL}, the $\indexED$th \ac{ED} calculates the numerals $\{\digit[\indexED,\indexGradient][\indexDigit]|\indexGradient\in\integers_\numberOfModelParameters, \indexDigit\in\integers_\numberOfDigits\}$ with \eqref{eq:encodedSequence}, for a given $\base$. The key strategy exploited at  the $\indexED$th \ac{ED} with the proposed scheme is that {\em$\base-1$ subcarriers are dedicated for each numeral  and one of these subcarriers is  activated  based on its value}. To express this encoding operation rigorously, let $\mappingFunction$ be a function that maps $\indexGradient\in\integers_\numberOfModelParameters$ to a set of $(\base-1)\numberOfDigits$ distinct time-frequency index pairs denoted by $\resourceSet[\indexGradient]\triangleq\{(\voteInTime[\indexDigit][\indexActiveSymbol],\voteInFrequency[\indexDigit][\indexActiveSymbol])|\indexDigit\in\integers_\numberOfDigits,\indexActiveSymbol\in\integers_{\base-1}\}$ for $\voteInTime[\indexDigit][\indexActiveSymbol]\in\integers_\numberOfOFDMSymbols$ and $\voteInFrequency[\indexDigit][\indexActiveSymbol]\in\integers_\numberOfActiveSubcarriers$, where $\resourceSet[\indexGradient_1]\cap\resourceSet[\indexGradient_2]=\emptyset$ if $\indexGradient_1\neq\indexGradient_2$ for $\indexGradient_1,\indexGradient_2\in\integers_\numberOfModelParameters$.
 %$\integersSkip_\base\triangleq\{0,1,\mydots,(\base-1)/2-1,(\base-1)/2+1,\mydots,\base-1\}\}$ 
 %Also, let $\transmittedSymbolAtSubcarrier[\indexED,{\voteInFrequency[+][\indexDigit]},{\voteInTime[+][\indexDigit]}]$ and $\transmittedSymbolAtSubcarrier[\indexED,{\voteInFrequency[-][\indexDigit]},{\voteInTime[-][\indexDigit]}]$ be two symbols dedicated to $\digit[\indexED,\indexGradient][\indexDigit]$.
%If the $\encoder$ is based on binary with sign, based on the value of $\digit[\indexED,\indexGradient][\indexDigit]$,  
 The $\indexED$th \ac{ED} determines the modulation symbol $\transmittedSymbolAtSubcarrier[\indexED,{\voteInTime[\indexDigit][\indexActiveSymbol],\voteInFrequency[\indexDigit][\indexActiveSymbol]}]$  as
\begin{align}
	\transmittedSymbolAtSubcarrier[\indexED,{\voteInTime[\indexDigit][\indexActiveSymbol],\voteInFrequency[\indexDigit][\indexActiveSymbol]}]=
	\sqrt{\symbolEnergy} \randomSymbolAtSubcarrier[\indexED,\indexGradient,\indexDigit]\times\indicatorFunction[{\digit[\indexED,\indexGradient][\indexDigit]=\symbol[\indexActiveSymbol]}]~,
	\label{eq:symbolPlus}
\end{align}
for all $\indexDigit\in\integers_\numberOfDigits$ and $\indexActiveSymbol\in\integers_{\base-1}$, where $\symbolEnergy=\base-1$ is a factor to normalize the OFDM symbol energy and $\randomSymbolAtSubcarrier[\indexED,\indexGradient,\indexDigit]$ is a randomization symbol on the unit circle for \ac{PMEPR} reduction \cite{sahinCommnet_2021}. Note that we do not allocate a subcarrier for $\symbol[\base-1]=0$ since it does not contribute to the sum given in \eqref{eq:digitAveraging}. After the calculation of \eqref{eq:symbolPlus} for all gradients, the $\indexED$th \ac{ED} calculates the \ac{OFDM} symbols and all \acp{ED} transmit them simultaneously based on the discussions in Section~\ref{sec:system}. Since the proposed scheme uses $(\base-1)\numberOfDigits$ subcarriers for each gradient, the maximum number of gradients that can be transmitted on each \ac{OFDM} symbol can be calculated as $\numberOfParametersPerOFDM=\floor{\numberOfActiveSubcarriers/((\base-1)\numberOfDigits)}$ for all \acp{ED}. 

 It is worth emphasizing that we do not use \ac{TCI} to compensate the impact of multipath channel on the transmitted symbols as this is beneficial to eliminate 1) the need for precise time synchronization, 2) the channel estimation overhead in a mobile wireless networks, 3) the information loss due to the truncation, and 4) the power instabilities in fading channel due to the channel inversion. %Our scheme also relies on a non-coherent receiver as discussed in Section~\ref{subsec:edrec}. 

\subsection{Edge Server - Receiver}
\label{subsec:edrec}
 At the \ac{ES},  we  assume that the \ac{CSI}, i.e., $\{\channelAtSubcarrier[\indexED,\indexSubcarrier,\indexOFDMSymbol]|\indexED\in\integers_\numberOfEdgeDevices,\indexSubcarrier\in\integers_\numberOfActiveSubcarriers,\indexOFDMSymbol\in\integers_\numberOfOFDMSymbols\}$, is {\em not} available. Hence, the \ac{ES} exploits that
 %due to the independent channel realizations between \acp{ED} and \ac{ES}, 
 $\receivedSymbolAtSubcarrier[{\voteInTime[\indexDigit][\indexActiveSymbol],\voteInFrequency[\indexDigit][\indexActiveSymbol]}]$ is a random vector for $\receivedSymbolAtSubcarrier[{\voteInTime[\indexDigit][\indexActiveSymbol],\voteInFrequency[\indexDigit][\indexActiveSymbol]}]\sim\complexGaussian[\zeroVector[\numberOfAntennasAtES]][{(\symbolEnergy\numberOFEDsForOptionGeneral[\indexActiveSymbol]+\noiseVariance)}{\identityMatrix[\numberOfAntennasAtES]}]$ and obtains an estimate of  $\{\numberOFEDsForOptionGeneral[\indexActiveSymbol]|\indexActiveSymbol\in\integers_{\base-1}\}$ to realize the last expression in \eqref{eq:digitAveraging}, non-coherently. For given $\indexDigit$ and $\indexGradient$, by using the corresponding log-likelihood function, the  \ac{ML} detector   can be expressed   as
\def\covarianceMatrix{{\bf \Sigma}_{\indexActiveSymbol}}
\def\aVector{{\rm \textbf{x}}_{\indexActiveSymbol}}
\begin{align}
	&\{\numberOFEDsForOptionGeneralDetector[\indexActiveSymbol]|\indexActiveSymbol\in\integers_{\base-1}\}
	=\arg\min_{\{\numberOFEDsForOptionGeneralVar[\forall\indexActiveSymbol]\}}\left\{\sum_{\indexActiveSymbol=0}^{\base-2}\ln\det\covarianceMatrix+\aVector^{\rm H}\covarianceMatrix^{-1}\aVector\right\}	\label{eq:originalProblem}
	\\&
	~~~~~~~~~~~~\text{s.t.}~  \numberOFEDsForOptionGeneralVar[\indexActiveSymbol]\in\{0,\mydots,\numberOfEdgeDevices\}, \indexActiveSymbol\in\integers_{\base-1}~, \sum_{\indexActiveSymbol=0}^{\base-2}\numberOFEDsForOptionGeneralVar[\indexActiveSymbol]\le\numberOfEdgeDevices~,
\nonumber
\end{align}
where $\aVector = [\Re\{\receivedSymbolAtSubcarrier[{\voteInTime[\indexDigit][\indexActiveSymbol],\voteInFrequency[\indexDigit][\indexActiveSymbol]}]\}^{\rm T} ~\Im\{\receivedSymbolAtSubcarrier[{\voteInTime[\indexDigit][\indexActiveSymbol],\voteInFrequency[\indexDigit][\indexActiveSymbol]}]\}^{\rm T}]^{\rm T}$ and $\covarianceMatrix=\frac{\symbolEnergy\numberOFEDsForOptionGeneralVar[\indexActiveSymbol]+\noiseVariance}{2}\identityMatrix[2\numberOfAntennasAtES]$. However, due to the constraints in \eqref{eq:originalProblem}, the complexity for a solution to \eqref{eq:originalProblem} at the receiver can be very high. To address this issue, we relax the constraints and evaluate $\numberOFEDsForOptionGeneralDetector[\indexActiveSymbol]$ independently as given by
\begin{align}
	\numberOFEDsForOptionGeneralDetector[\indexActiveSymbol]&=\arg \min_{\numberOFEDsForOptionGeneralVar[\indexActiveSymbol]} \left\{2\numberOfAntennasAtES\ln\left(\frac{\symbolEnergy\numberOFEDsForOptionGeneralVar[\indexActiveSymbol]+\noiseVariance}{2}\right)+\frac{2\norm{\receivedSymbolAtSubcarrier[{\voteInTime[\indexDigit][\indexActiveSymbol],\voteInFrequency[\indexDigit][\indexActiveSymbol]}]}_2^2}{\symbolEnergy\numberOFEDsForOptionGeneralVar[\indexActiveSymbol]+\noiseVariance}\right\}\nonumber\\
	&=\frac{\norm{\receivedSymbolAtSubcarrier[{\voteInTime[\indexDigit][\indexActiveSymbol],\voteInFrequency[\indexDigit][\indexActiveSymbol]}]}_2^2}{\symbolEnergy\numberOfAntennasAtES}-\frac{\noiseVariance}{\symbolEnergy}~.
	\label{eq:edEst}
\end{align}
Thus, a low-complexity estimator of $\digitAveraged[\indexGradient][\indexDigit]$   can be obtained as
\begin{align}
	\digitAveragedEst[\indexGradient][\indexDigit]=
	\frac{1}{\numberOfEdgeDevices}\sum_{\indexActiveSymbol=0}^{\base-2} \symbol[\indexActiveSymbol]\numberOFEDsForOptionGeneralDetector[\indexActiveSymbol]~.
	\label{eq:digitAveEst}
\end{align}
Finally, the estimator of $\meanGradientEle[\indexCommunicationRound][\indexGradient]$ can be expressed as
\begin{align}
	\meanGradientEleEstimate[\indexCommunicationRound][\indexGradient]=\decoder\representationInBase[\base][{\digitAveragedEst[\indexGradient][\numberOfDigits-1],\mydots,\digitAveragedEst[\indexGradient][1],\digitAveragedEst[\indexGradient][0]}]~.
	\label{eq:finalValEst}	
\end{align}
The \ac{ES} then transmits $\meanGradientVectorEstimate[\indexCommunicationRound]$ to the \acp{ED} for the next communication round and the $\indexED$th \ac{ED} updates its parameters as $
	\modelParametersAtIteration[\indexCommunicationRound+1] = \modelParametersAtIteration[\indexCommunicationRound] - \learningRate  \meanGradientVectorEstimate[\indexCommunicationRound]
$, $\forall\indexED$.

%Note that if the \ac{ES} does not know the number of \ac{ED} participating into training in advance (e.g. grantless access), inclusion a subcarrier for the zero-valued numerals can be helpful for the estimation of \eqref{eq:digitAveraging} can be estimated as $\numberOfEdgeDevices=$  based on \eqref{eq:edEst} and \eqref{eq:digitAveEst} 

\begin{figure*}[t]
	\centering
	{\includegraphics[width =\textwidth-20mm]{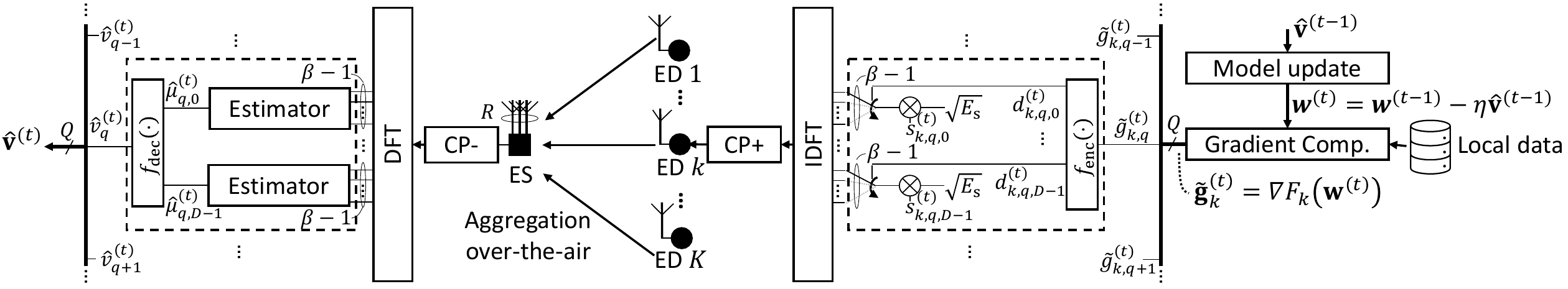}
	} 
	\caption{The transmitter and receiver diagrams with the proposed \ac{OAC} scheme for \ac{FEEL}.} 
	\label{fig:feelBlockDiagram}
\end{figure*}

The transmitter and received diagrams based on the aforementioned discussions are provided in \figurename~\ref{fig:feelBlockDiagram}.

\subsection{MSE Analysis}
\label{subsec:mse}
The variable $\norm{\receivedSymbolAtSubcarrier[{\voteInTime[\indexDigit][\indexActiveSymbol],\voteInFrequency[\indexDigit][\indexActiveSymbol]}]}_2^2/\numberOfAntennasAtES$ in \eqref{eq:edEst} is the average of $\numberOfAntennasAtES$ exponential variables with the mean  $\symbolEnergy\numberOFEDsForOptionGeneral[\indexActiveSymbol]+\noiseVariance$.  Thus,
the distribution of $\norm{\receivedSymbolAtSubcarrier[{\voteInTime[\indexDigit][\indexActiveSymbol],\voteInFrequency[\indexDigit][\indexActiveSymbol]}]}_2^2/\numberOfAntennasAtES$ is  $\gammaDist[\numberOfAntennasAtES][{\numberOfAntennasAtES/(\symbolEnergy\numberOFEDsForOptionGeneral[\indexActiveSymbol]+\noiseVariance)}]$. 
As a result, the mean and the variance of the estimator $\numberOFEDsForOptionGeneralDetector[\indexActiveSymbol]$ can be calculated via the properties of a gamma distribution as
\begin{align}
\expectationOperator[{\numberOFEDsForOptionGeneralDetector[\indexActiveSymbol]}][{}]=\frac{\expectationOperator[\norm{\receivedSymbolAtSubcarrier[{\voteInTime[\indexDigit][\indexActiveSymbol],\voteInFrequency[\indexDigit][\indexActiveSymbol]}]}_2^2/\numberOfAntennasAtES][]}{\symbolEnergy}-\frac{\noiseVariance}{\symbolEnergy}=\numberOFEDsForOptionGeneral[\indexActiveSymbol]~,
\end{align}
and
\begin{align}
\varianceOperator[{\numberOFEDsForOptionGeneralDetector[\indexActiveSymbol]}]=\frac{\varianceOperator[\norm{\receivedSymbolAtSubcarrier[{\voteInTime[\indexDigit][\indexActiveSymbol],\voteInFrequency[\indexDigit][\indexActiveSymbol]}]}_2^2/\numberOfAntennasAtES]}{\symbolEnergy}=\frac{1}{\numberOfAntennasAtES}\left(\numberOFEDsForOptionGeneral[\indexActiveSymbol]+\frac{\noiseVariance}{\symbolEnergy}\right)^2~,
\label{eq:variance}
\end{align}
respectively, where the expectation is calculated over the randomness of the channel and noise. Hence, $\numberOFEDsForOptionGeneralDetector[\indexActiveSymbol]$ is an unbiased estimator. Also, based on \eqref{eq:digitAveEst} and \eqref{eq:finalValEst}, both $\digitAveragedEst[\indexGradient][\indexDigit]$ and $\meanGradientEleEstimate[\indexCommunicationRound][\indexGradient]$ are unbiased estimators of $\digitAveraged[\indexGradient][\indexDigit]$ and $\meanGradientEleOverQuantized[\indexCommunicationRound][\indexGradient]$, respectively. For a given  $\{\numberOFEDsForOptionGeneral[\indexActiveSymbol]| \indexActiveSymbol\in\integers_{\base-1}\}$, by using \eqref{eq:digitAveEst} and \eqref{eq:variance},  the variance of the estimator $\digitAveragedEst[\indexGradient][\indexDigit]$ is obtained as
\begin{align}
\varianceOperator[{{\digitAveragedEst[\indexGradient][\indexDigit]}}]=\frac{1}{\numberOfAntennasAtES\numberOfEdgeDevices^2}\sum_{\indexActiveSymbol=0}^{\base-2}\symbol[\indexActiveSymbol]^2\left(\numberOFEDsForOptionGeneral[\indexActiveSymbol]+\frac{\noiseVariance}{\symbolEnergy}\right)^2~.
\end{align} 
Thus, we can calculate the variance of the estimator ${{\meanGradientEleEstimate[\indexCommunicationRound][\indexGradient]}}$ as
\begin{align}
	&\varianceOperator[{{\meanGradientEleEstimate[\indexCommunicationRound][\indexGradient]}}]=\frac{\valueMaximum^2}{\normalizationDigit^2\numberOfAntennasAtES\numberOfEdgeDevices^2}\sum_{\indexDigit=0}^{\numberOfDigits-1}\sum_{\indexActiveSymbol=0}^{\base-2}\symbol[\indexActiveSymbol]^2\left(\numberOFEDsForOptionGeneral[\indexActiveSymbol]+\frac{\noiseVariance}{\symbolEnergy}\right)^2\base^{2\indexDigit}~.
	\label{eq:varGivenK}
\end{align}
Hence, the \ac{MSE} of the estimator  $\meanGradientEleEstimate[\indexCommunicationRound][\indexGradient]$  can be obtained as
\begin{align}
	&\classicalMSEOperator[{\meanGradientEleEstimate[\indexCommunicationRound][\indexGradient]}]
	=\nonumber\\&\frac{\valueMaximum^2}{\normalizationDigit^2\numberOfAntennasAtES\numberOfEdgeDevices^2}\sum_{\indexDigit=0}^{\numberOfDigits-1}\sum_{\indexActiveSymbol=0}^{\base-2}\symbol[\indexActiveSymbol]^2\left(\numberOFEDsForOptionGeneral[\indexActiveSymbol]+\frac{\noiseVariance}{\symbolEnergy}\right)^2\base^{2\indexDigit}+\underbrace{\frac{1}{\numberOfEdgeDevices^2}\left(\sum_{\indexED=0}^{\numberOfEdgeDevices-1} \localGradientElementQuantized[\indexED,\indexGradient][\indexCommunicationRound]-\localGradientElement[\indexED,\indexGradient][\indexCommunicationRound]\right)^2}_{(\meanGradientEle[\indexCommunicationRound][\indexGradient]-\meanGradientEleOverQuantized[\indexCommunicationRound][\indexGradient])^2}\nonumber~,
\end{align}	
where the last term  is the squared bias  due to the quantization error. From the MSE, we can infer that the error  rapidly diminishes either by increasing $\base$ and $\numberOfDigits$, at a cost of increased number of resources, or increasing $\numberOfEdgeDevices$.

\def\metric[#1][#2]{m_{#1}^{(#2)}}
\def\metricVector[#1]{\textbf{m}^{(#1)}}
\def\metricFactor{\alpha}

\section{Numerical Results}
\label{sec:numerical}

To numerically analyze \ac{OAC} with the proposed scheme for \ac{FEEL}, we consider the learning task of handwritten-digit recognition in a single cell with $\numberOfEdgeDevices=25$ \acp{ED}. We set the \ac{SNR}, i.e., $1/\noiseVariance$, to be $20$~dB, and choose the number of antennas at the \ac{ES} as $\numberOfAntennasAtES\in\{1,25\}$. 
For the fading channel, we consider ITU Extended Pedestrian A (EPA) with no mobility and regenerate the channels between the \ac{ES} and the \acp{ED} independently for each communication round to capture the long-term channel variations.
The subcarrier spacing is set to $15$~kHz. We use  $\numberOfActiveSubcarriers=1200$ subcarriers (i.e., the signal bandwidth is $18$~MHz). Hence, the difference between time of arriving \ac{ED} signals is maximum $\syncError=55.6$~ns. We assume that the synchronization uncertainty at the \ac{ES} is $\Nerror=3$ samples. For the comparisons, we consider \ac{FSK-MV} proposed in \cite{sahinCommnet_2021} as it is based on a non-coherent detection and provides robustness against time-synchronization errors. We do not consider methods based on precoding such as \ac{TCI} since their performance can deteriorate quickly in the cases of time-synchronization errors \cite{sahinWCNC_2022,sahinCommnet_2021} or imperfect \ac{CSI} \cite{Haejoon_2021}. We consider $\base\in\{3,5,7\}$ and $\numberOfDigits=\{1,2\}$.

For the local data at the \acp{ED}, we use the MNIST database that contains labeled handwritten-digit images size of $28\times28$ from digit 0 to digit 9. We distribute the data samples in the MNIST database to the \acp{ED} to generate representative results for \ac{FEEL}. We consider both homogeneous and heterogeneous data distributions in the cell. To prepare the data, we first choose $|\completeData|=25000$ training images from the database, where each digit has distinct $2500$ images.  
For the scenario with the homogeneous data distribution, we assume that each \ac{ED} has $250$ distinct images for each digit. As done in \cite{sahinCommnet_2021}, for the scenario with  the heterogeneous data distribution, we divide the cell into 5 areas with concentric circles and the \acp{ED} located in $\indexArea$th area have the data samples with the labels $\{\indexArea-1,\indexArea,1+\indexArea,2+\indexArea,3+\indexArea,4+\indexArea\}$ for $\indexArea\in\{1,\mydots,5\}$ (See \cite[Figure 3]{sahinCommnet_2021} for an illustration). The number of \acp{ED} in each area is $5$. As discussed in Section~\ref{sec:system}, we assume that the path loss is compensated through a power control mechanism. For the model, we consider a \ac{CNN}  given  in \cite[Table~1]{sahinCommnet_2021}. At the input layer, standard normalization is applied to the data. Our model has $\numberOfModelParameters=123090$ learnable parameters.  For the update rule, the learning rate is set to $0.001$. The batch size $\batchSize$ is $64$. For the proposed scheme, we use \ac{SGD} with the momentum $0.9$. For the test accuracy, we use $10000$ test samples available in the MNIST database.

\def\figuresize{\textwidth/2-15mm}
\begin{figure}
	\centering
	\subfloat[$\numberOfAntennasAtES=1$.]{\includegraphics[width =\figuresize]{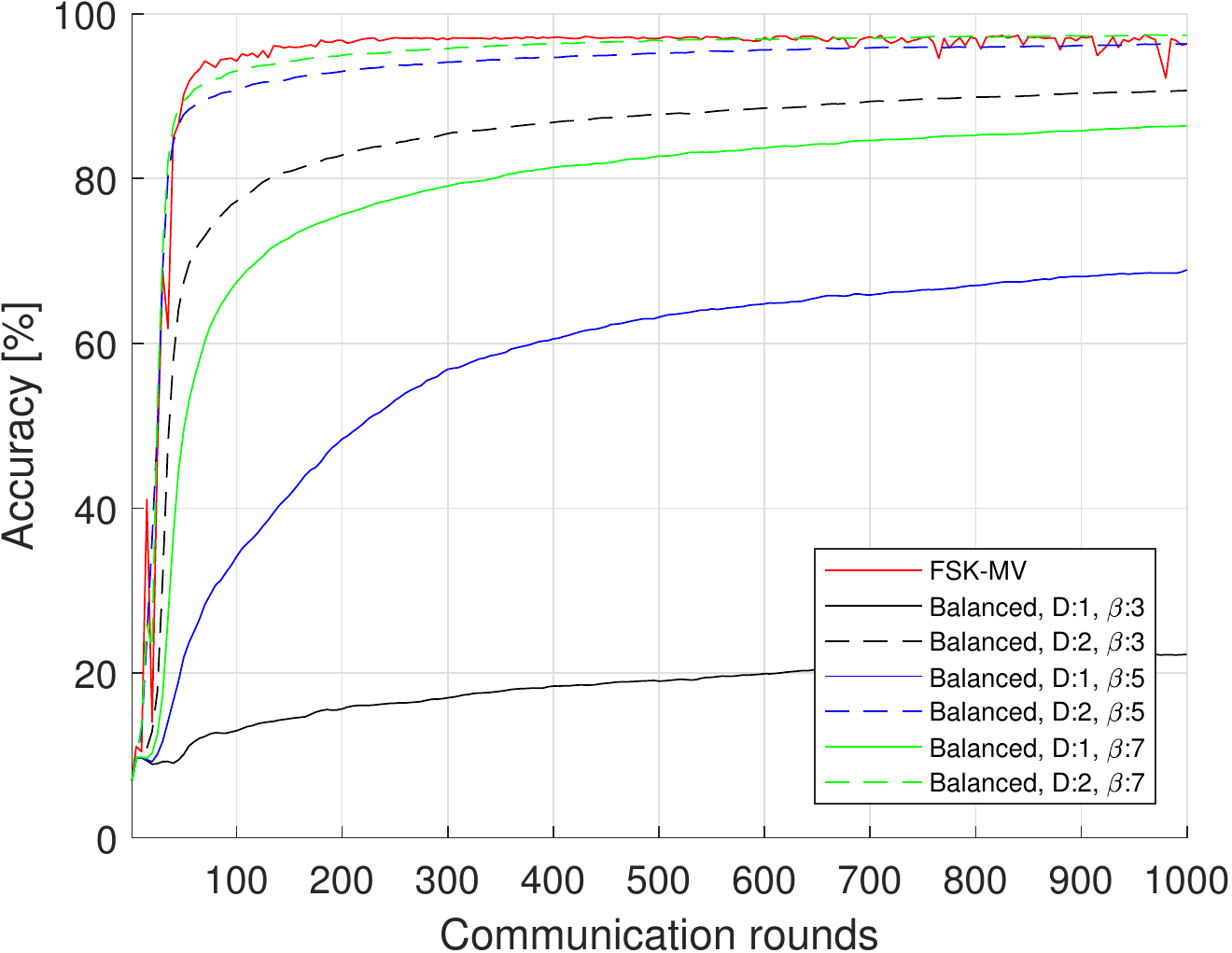}
	\label{subfig:acc_homo_R1_wM_woAAM}}	
\\	
	\subfloat[$\numberOfAntennasAtES=25$.]{\includegraphics[width =\figuresize]{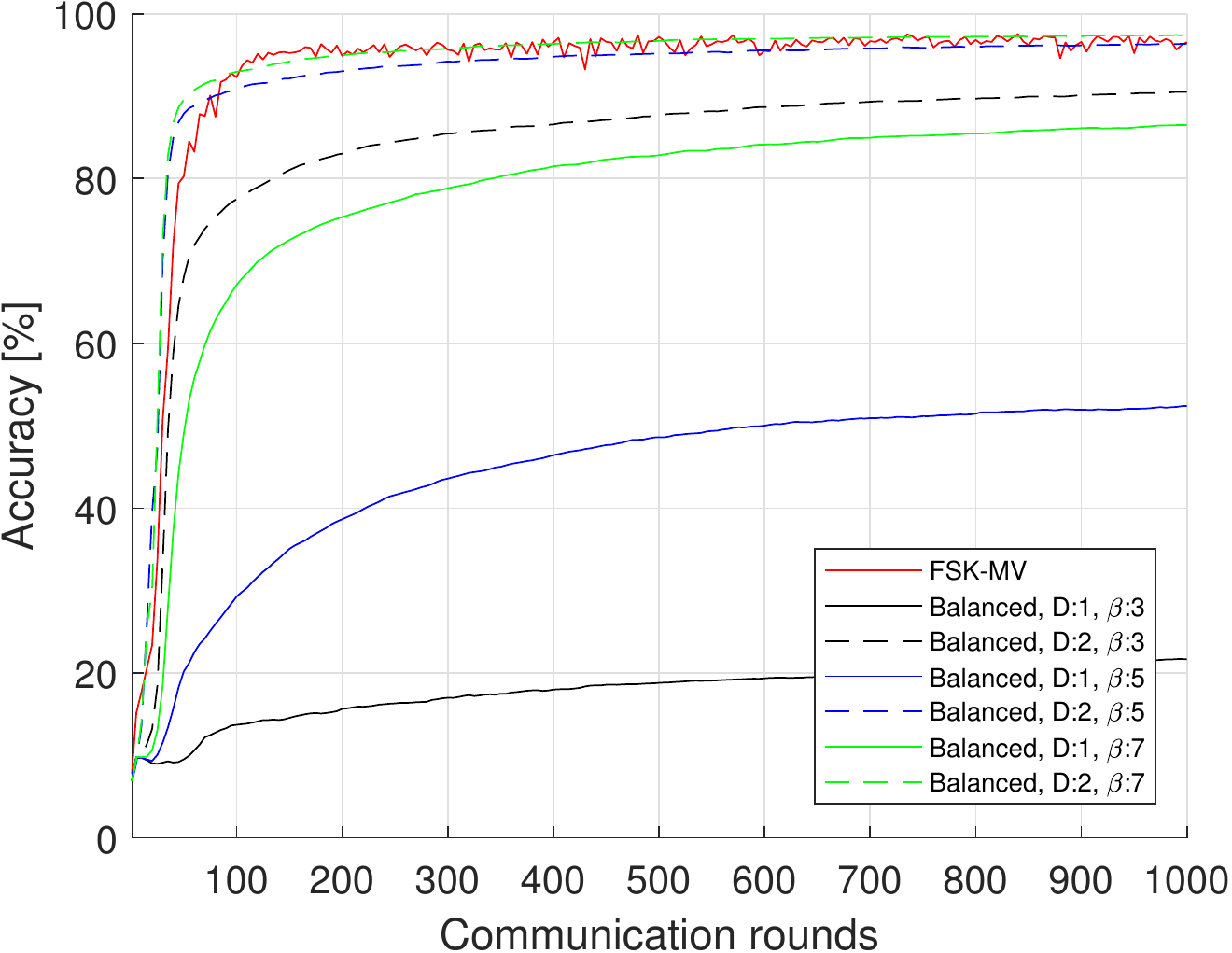}
	\label{subfig:acc_homo_R25_wM_woAAM}}					
	\caption{Test accuracy (Homogeneous data distribution).}
	\label{fig:testAccHomo}
\end{figure}
In \figurename~\ref{fig:testAccHomo}, we provide the test accuracy versus communication rounds for the scenario with homogeneous data distribution.  In \figurename~\ref{fig:testAccHomo}\subref{subfig:acc_homo_R1_wM_woAAM}, there is only a single antenna at the \ac{ES}. Although the accuracy results with the proposed scheme improve for larger $\base$ or $\numberOfDigits$ (i.e., less $\coefficientOne$) for this scenario, the \ac{FSK-MV} is superior to the proposed scheme in terms of convergence rate. This is because \ac{FSK-MV} is based on \ac{signSGD}, while the proposed scheme implements \ac{SGD}. In \cite{Bernstein_2018}, it was also mentioned that \ac{signSGD} can outperform \ac{SGD} by providing stronger weight to the gradient direction as compared to \ac{SGD} when the gradients are noisy.  In \figurename~\ref{fig:testAccHomo}\subref{subfig:acc_homo_R25_wM_woAAM},  we set the number of antennas at the \ac{ES} to be $25$. Although this improves the \ac{MSE} considerably, its impact on the test accuracy is almost negligible. Hence, the proposed scheme can achieve notable test accuracy results even when there is only a {\em single} antenna at the \ac{ES}.

\begin{figure}
	\centering
%	\subfloat[Without AAM (Momentum: 0, $\numberOfAntennasAtES=1$).]{\includegraphics[width =\figuresize]{figure_acc_learningRate0.001_momentum0_aam_0_Nantenna1_synch1_NxED_25_Areas_5_ND_25000-eps-converted-to.pdf}
%		\label{subfig:acc_heto_R1_woM_woAAM}}	
	\subfloat[$\numberOfAntennasAtES=1$.]{\includegraphics[width =\figuresize]{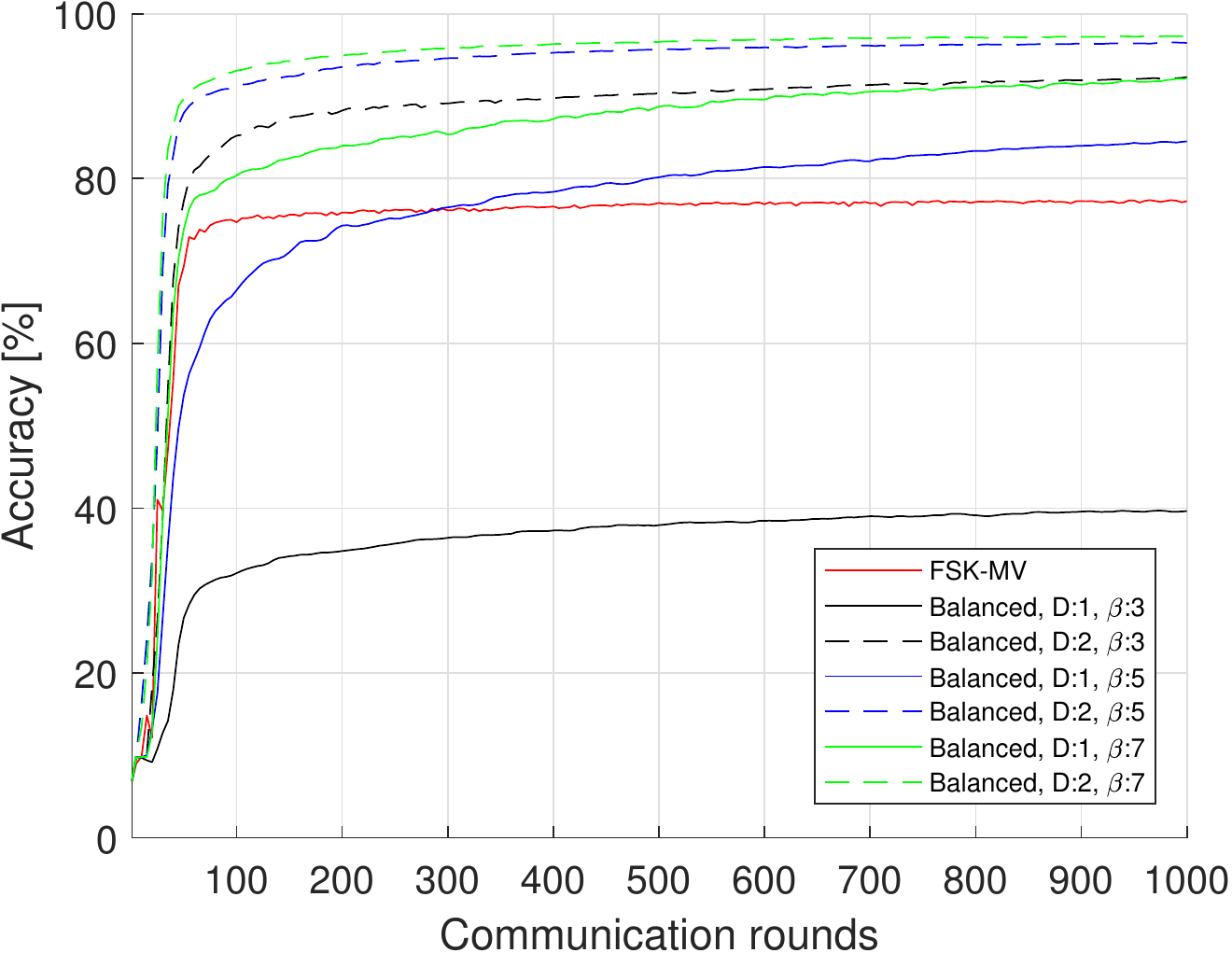}
	\label{subfig:acc_heto_R1_wM_woAAM}}		
\\
%	\subfloat[Without AAM (Momentum: 0, $\numberOfAntennasAtES=25$).]{\includegraphics[width =\figuresize]{figure_acc_learningRate0.001_momentum0_aam_0_Nantenna25_synch1_NxED_25_Areas_5_ND_25000-eps-converted-to.pdf}
%		\label{subfig:acc_heto_R25_woM_woAAM}}			
	\subfloat[$\numberOfAntennasAtES=25$.]{\includegraphics[width =\figuresize]{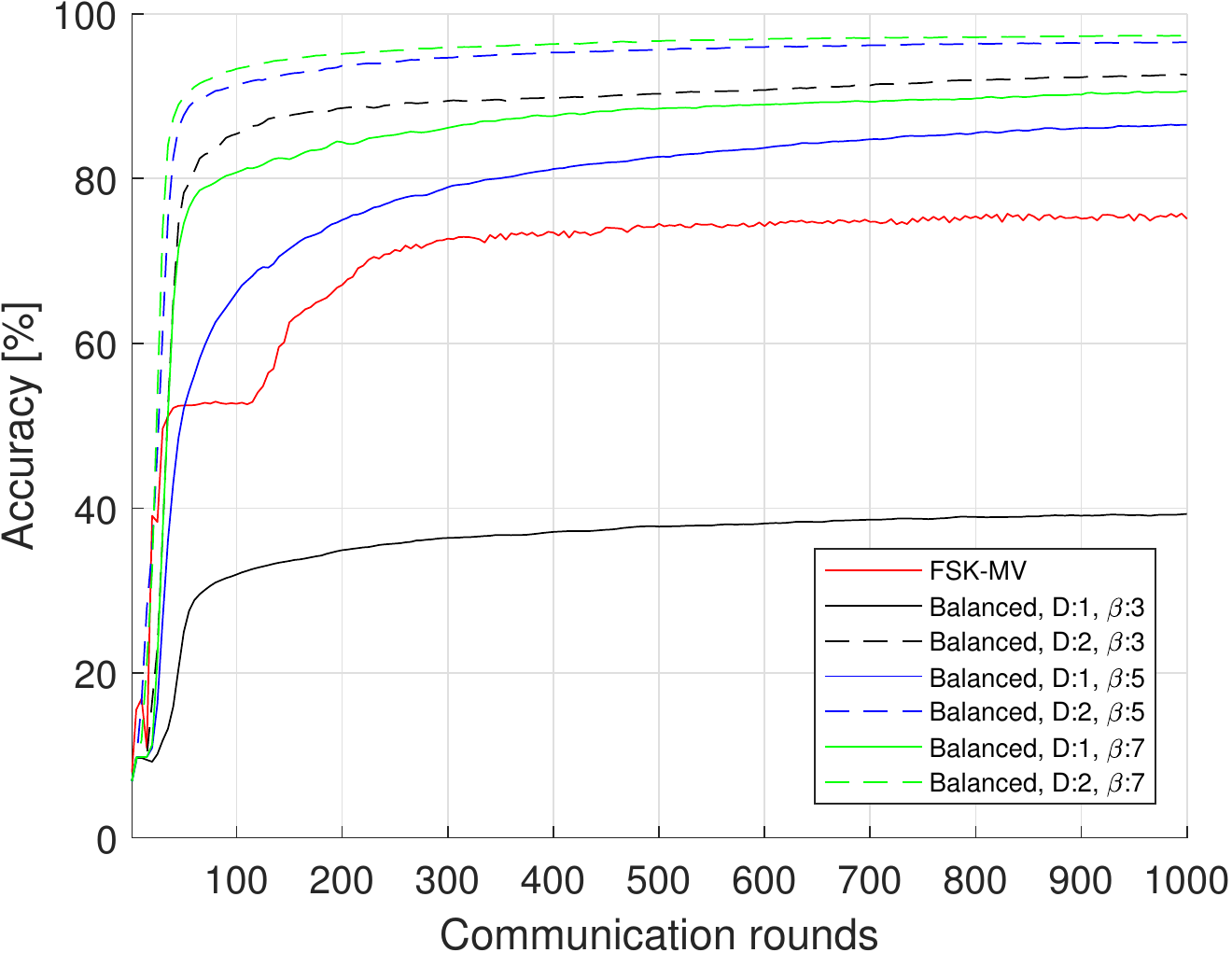}
	\label{subfig:acc_heto_R25_wM_woAAM}}		
	\caption{Test accuracy (Heterogeneous data distribution).}
	\label{fig:testAccHete}
\end{figure}

In \figurename~\ref{fig:testAccHete}, the test accuracy is evaluated when the data distribution is highly heterogeneous, i.e., each \ac{ED} has only 6 unique digits. We use the same parameters used for \figurename~\ref{fig:testAccHomo}. In this case, the performance of the \ac{FSK-MV}  degrades drastically, whereas the performance of the proposed scheme is similar to one in \figurename~\ref{fig:testAccHomo}. The test accuracy under heterogeneous data distribution is less than 80\% for the \ac{FSK-MV} (this is also reported in \cite{sahinCommnet_2021}). On the other hand, the proposed scheme with a larger set of $\base$ and $\numberOfDigits$ can achieve more than 90\% test accuracy as shown in \figurename~\ref{fig:testAccHete}\subref{subfig:acc_heto_R1_wM_woAAM} for $\numberOfAntennasAtES=1$. A similar observation can also be made for $\numberOfAntennasAtES=25$ as in \figurename~\ref{fig:testAccHete}\subref{subfig:acc_heto_R25_wM_woAAM}, i.e., it can provide high accuracy, up to 98\%, even when the data distribution is not homogeneous.

\section{Concluding Remarks}
\label{sec:conclusion}

In this study, we investigate an \ac{OAC} method that exploits balanced number systems for gradient aggregation. The proposed scheme achieves a continuous-valued computation through a digital scheme by exploiting the fact that the average of the numerals in the real domain can be used to compute the average of the corresponding real-valued parameters approximately. With the proposed \ac{OAC} method, the local stochastic gradients are encoded into a sequence where the elements of the sequence determine the activated \ac{OFDM} subcarriers. We also use a non-coherent receiver to eliminate the precise sample-level time synchronization and channel estimation overhead due to the channel inversion techniques. We show that the proposed scheme results in a lower MSE at the expense of more resources. Future work will present the corresponding convergence analysis.
% omitted in the study due to the page limitation.

 %To improve its MSE performance, we also introduce \ac{AAM}. We theoretically analyze its MSE performance and its convergence rate for \ac{FEEL} that consider both homogeneous and heterogeneous distributions. Our numerical results demonstrate that the test accuracy of the \ac{FEEL} with the proposed scheme using \ac{AAM} can reach up to 98\% even when the \acp{ED} do not have the labels in their data sets. 

%The proposed scheme provides a potentially rich area to be investigated. For example, in this study, we consider gradient aggregation. On the other hand, one open question is if the proposed scheme can also be utilized for parameter aggregation. Based on our numerical tests, the performance (e.g., test accuracy) can be poor as the neural network may not be tolerant to the errors on the model parameters due to the proposed scheme. Hence, evaluating (and enhancing) the proposed scheme with a noise-tolerant neural network (e.g., quantized neural networks) is an interesting future research direction that can be pursued. Another interesting direction is the utilization of the proposed \ac{OAC} scheme along with distributed source coding to reduce the per-round communication latency further.

\acresetall

\bibliographystyle{IEEEtran}
\bibliography{references}

\end{document}